\documentclass[12pt]{article}
\usepackage{amsfonts}
\usepackage{amsmath}
\usepackage[labelfont=bf]{caption}
\usepackage{float}
\usepackage{footmisc}
\usepackage{graphicx,datetime}
\usepackage{geometry}
\usepackage{hyperref}
\usepackage{mathtools}
\usepackage{natbib}
\usepackage{tikz}
\usepackage{setspace}
\usepackage{xcolor}

\usetikzlibrary{calc}

\newcommand*{\Resize}
[2]{\resizebox{\textwidth}{!}{$#2$}}
\newcommand\myeq{\mathrel{\stackrel{\makebox[0pt]{\mbox{\normalfont\tiny
					l'Hopital}}}{=}}}

\newtheorem{proposition}{Proposition}

\hypersetup{
	colorlinks,
	linkcolor={red!50!black},
	citecolor={blue!50!black},
	urlcolor={blue!80!black}
}

\begin{document}

	\title{Truly Costly Search and Word-of-Mouth Communication%
		\thanks{
			I thank Maarten Janssen for helpful suggestions and 
			comments.  Financial support from uni:docs 
			Fellowship Program is acknowledged.
		}
}
	\author{Atabek Atayev
		\thanks{%
			Corresponding author. E-mail: atabek.atayev@zew.de}
			\\
			\small{ZEW---Leibniz Centre for European 
				Economic Research in Mannheim}	
}
		\date{\today}
		
		\maketitle

\begin{abstract}
	\noindent In markets with search frictions, consumers can 
	acquire information about goods either through costly search 
	or from friends via word-of-mouth (WOM) communication.  How 
	do sellers' market power react to a very large increase 
	in the number of consumers' friends with whom they engage in 
	WOM? The answer to the question depends on 	whether 
	consumers are freely endowed with price information.  If 	
	acquiring price quotes is costly, equilibrium prices are 
	dispersed and the expected price is higher than the marginal 
	cost of production.  This implies that firms retain market 
	power even if price information is disseminated among a very 
	large number of consumers due to technological 
	progress, such as social networking websites.
	\\
	
	
\end{abstract}
\newpage

	\sloppy
\section{Introduction}

It is well documented that consumers can acquire information 
about goods not only on their own but also through their 
personal contacts, or friends, acquaintances.%
\footnote{Due to \cite{stigler1961}, most studies on consumer
	search focus on acquisition of information by consumers on 
	their own.  A pioneer study by \cite{katzlazarsfeld1955} 
	show that consumers' personal contacts are a key reason why 
	they buy goods. Later empirical studies back this point, 
	e.g. \cite{godesmayzlin2004}, \cite{chenetal2011}.}
Information dissemination through friends via word-of-mouth 
(WOM) communication is important due to recent technological 
growth, such as developments of social websites, online 
forums, and instant messengers.  Consumers can exchange 
information with a greater number of friends as technology 
progresses further.  Naturally, this puts forward a question, 
how do markets respond to an increase in the number of 
consumers' personal contacts?

Certainly, the answer to this question depends on many 
characteristics of a market under consideration.  To illustrate 
the relationship between consumers' number of friends and 
prices, the existing literature considers the following 
canonical environment. In a duopoly market, symmetric firms 
produce homogeneous goods and compete in prices.  Consumers do 
not know prices. They can acquire price quotes through costly 
search or may receive such information from friends, who have 
searched, for free. The existing literature shows that, if each 
consumer accesses a price quote of a random seller for free, 
prices converge to the sellers' marginal cost of production as 
the number of consumers' friends increases 
(\cite{galeotti2010}).  Yet, we show in this paper that if 
consumers are not endowed with any free price quote, prices are 
dispersed across sellers and the expected price is greater than 
the production marginal cost even when the number consumers' 
friends increases without limits.  

The intuition behind our main result is as follows.  Sellers' 
market power and the level of price dispersion are determined by 
the share of consumers who only observe one price, also known as 
\textit{locked-in} or \textit{captive} consumers 
(e.g., \cite{shilony1977}, \cite{varian1980}, 
\cite{armstrongvickers2019}).  Since captive consumers cannot 
``walk away'' to a competing seller, sellers' market power 
increases with the share of captive consumers. Price dispersion 
exists as long as the share of captive consumers does not vanish 
or not all consumers become captive.  An increase in the number 
of consumers' friends has two opposing impacts on the share of 
captive consumers.  First is the direct competitive effect.  
Search intensity of consumers being fixed, the share of captive 
consumers falls with the number of friends because some of the 
captive consumers now have friends from whom they can receive 
another price quote.  This puts competitive pressure on firms.  
There is also an indirect effect, which is anti-competitive.  
The share of consumers, who search and acquire price quotes 
through costly search, decreases with the number of 
friends.  This is a standard free-riding effect. As information 
on prices is a public good for consumers, an increase in the 
number of personal contacts leads to a lower level of 
information acquisition.  As a result less information diffuses 
throughout consumers.  This leads to a higher share of captive 
consumers and, thus, raises sellers' market power.

In an extreme case when the number of consumers' personal 
contacts increases without bounds, the two effects discussed in 
the previous paragraph interact in a way so that sellers retain 
some market power.  The reason is that, if searching for prices 
is costly, the share of consumers who search vanishes as the 
number of consumers' friends gets very large.  Despite the fact 
that \textit{almost} no consumer searches, very little amount of 
price information is still obtained.  This information diffuses 
throughout a very large number of consumers so that there is a 
strictly positive share of captive consumers as well as that of 
price-comparing consumers.  As a result, price dispersion 
remains. Since sellers never price below the production marginal 
cost, the presence of price dispersion means that the expected 
price is higher than the production marginal cost.

Our result differs from that of \cite{galeotti2010} because, in 
\cite{galeotti2010} each buyer knows a price quote of a firm 
without searching.  This means that even if consumers do not 
search, they may still compare prices because they receive price 
information from friends.  Therefore, when the number of 
consumers' friends gets very high so that buyers stop searching, 
all consumers compare prices because, in addition to a free 
price quote a consumer observes, each buyer almost certainly 
receives one more price quote from friends.  If all 
consumers compare prices, firms find it optimal to price at the 
production marginal cost.

Our paper contributes to the growing literature on consumer 
search and WOM, and the closest papers to ours are 
\cite{atayevjanssen2019}, \cite{campbelletal2020}, and 
\cite{galeotti2004}).   \cite{atayevjanssen2019}, 
which unlike the current paper studies sequential search 
markets, shows that price dispersion persists when the number of 
consumers' personal contacts links increases. 
\cite{campbelletal2020} examines the relationship between 
network structure of consumers' personal contacts and product 
quality.  \cite{galeotti2004}, which is an earlier 
version of \cite{galeotti2010}, employs the same model as that 
in our current paper, but the author does not examine an impact 
of an (unbounded) increase in the number of buyers' personal 
contacts on market outcomes.  


The rest of our paper is organized as follows.  In the next 
section, we revisit \cite{galeotti2004} and examine the limiting 
case when the number of consumers' friends increases without 
bounds.  In Section \ref{s:link_form}, we show that our main 
result holds even when forming a personal link is costly and, 
thus, consumers have to strategically decide whether to form 
links.  The final section concludes.

\section{Model and Analysis}\label{s:simultaneous}

In this section, we analyze a model of \cite{galeotti2004}.  
There are two firms, or sellers, that produce homogeneous goods 
at a marginal cost normalized to zero. The firms compete in 
prices.  Let $F_j(p)$ represent the probability that firm $j$ 
charges a price that is not greater than $p$. 

On the demand side, there is a countably infinite number of 
consumers, normalized to one.  Each consumer demands a unit of 
the good, which she evaluates at $v>0$. Prior to searching, a 
consumer does not know prices and, to make a purchase, she must 
learn at least one price quote.  To do that, a buyer can search 
firms, which is costly.   Alternatively, a buyer can obtain 
price information from her friends at no cost, given her friends 
acquired price quotes themselves.  The cost of searching a firm 
is given by $c>0$.  Search is simultaneous, meaning that a 
consumer requests price quotes from $n\in \left\{0,1,2\right\}$ 
firms simultaneously (implying that when $n=0$ she just waits), 
after which search is terminated.   Let $q_n$ be the probability 
that a consumer samples $n$ price quotes: 
$\sum_{n=0}^{2}q_n=1$.  Each consumer is assumed to have $k$ 
number of friends, or personal links, who may potentially inform 
the consumer about prices, where $1\leq k < \infty$ .%
\footnote{As the paper is concerned with information diffusion
	among consumers, an uninteresting case of $k=0$ is omitted, 
	yet it can be easily incorporated.}

Timing of the game is as follows. First, firms simultaneously 
set prices.  Second, without knowing prices, consumers choose 
number of firms to search.  Third, after all consumers finish 
their search, they share price information with their friends.  
Consumers who learn at least a single price---either through own 
search or from friend(s)---may make purchase.  We employ 
Nash Equilibrium (NE) as a solution concept.

\cite{galeotti2004} shows that for sufficiently large $k$, there 
exists a stable NE where consumers randomize between 
searching one firm and not searching, i.e. $q_0, q_1>0$ and 
$q_0+q_1=1$.  Letting $q \equiv q_1$ for the rest of the paper 
so that $q_0=1-q$ and $\eta \equiv 
\frac{\left(1-\frac{q}{2}\right)^{k+1} - (1-q)^{k+1}}{1 + 
(1-q)^{k+1} - 2 \left(1-\frac{q}{2}\right)^{k+1}}$, we restate 
the result in the following proposition.
\begin{proposition}[\cite{galeotti2004} Theorem 5.1]\label{prop:simul_eq}
	For any $v>0$ and $k\geq 1$, there exists 
	$[\underline{c}(k),\overline{c}]\subset (0,v)$ such that for 
	any $c \in 	[\underline{c}(k), \overline{c}]$ there 
	exists a stable NE given by $(F(p), q)$ where
	\begin{equation}\label{eq:F_simul}
	F(p) = 1 - \eta \left(\frac{v}{p}-1\right), \ \mbox{with support} \ 
	\left[\frac{\eta}{1 +  
		\eta}v,v\right],
	\end{equation}
	 and $q$ solves
	 \begin{equation}\label{eq:IC_simul}
	 c = (1-q)^k(v-E[p]) + \left(\left(1-\frac{q}{2}\right)^k - 
	 (1-q)^k\right)\big(E[p] - E[\min\{p_1,p_2\}]\big).
	 \end{equation}
\end{proposition}

We refer to the appendix of \cite{galeotti2004} for the proof.  
The intuition is as follows.  First, the author assumes that 
buyers randomize between searching one firm and not searching.  
Given this search strategy, he determines sellers' pricing 
policies, which is given in \eqref{eq:F_simul}.  Note that, for 
any strictly positive shares of searching consumers and 
non-searching consumers, the price distribution is 
non-degenerate, meaning that prices are dispersed.  Second, the 
author verifies whether, given such dispersed prices and buyers' 
number of friends, it is optimal for buyers randomize between 
searching one firm and not searching.  He shows that this is 
indeed the case if the search cost is not too high and not too 
low.  Moreover, the authors shows that there exists either one 
or two such equilibria, but only one of them is stable.  

\begin{figure}[h]
	\centering
	\begin{tikzpicture}
		\node[inner sep=0pt]  at (0,0)
		{\includegraphics[width=.6\linewidth]{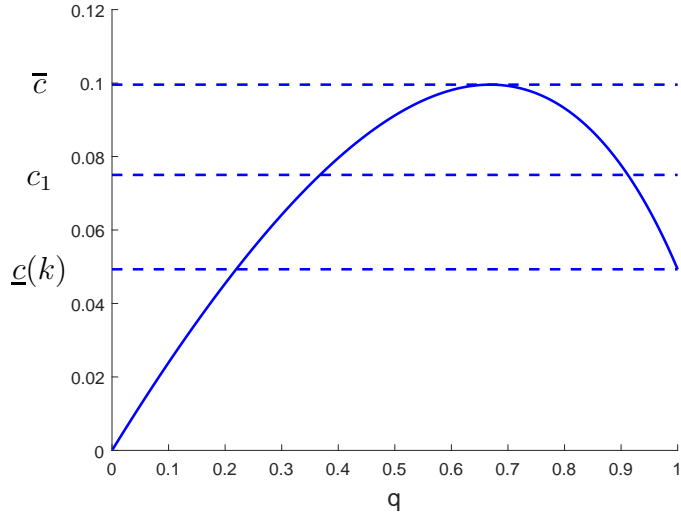}};
		\node[inner sep=0pt] at (-4.5,-.4) {$\underline{c}(k)$};
		\node[inner sep=0pt] at (-4.5,.8) {$c_1$};
		\node[inner sep=0pt] at (-4.5,2.1) {$\overline{c}$};
	\end{tikzpicture}
	\caption{Illustration of equilibria existence
	$v=1$, $k=1$, and $c_1 = 0.075$}
	\label{fig:equil_q1}
\end{figure}

To illustrate the main idea, we present Figure 
\ref{fig:equil_q1}.  The horizontal axis represents the share of 
consumers who search one firm, i.e., $q$, and the vertical axis 
stands for the expected benefit as well as the cost of searching 
a firm.  The solid curve represents the expected benefit of 
searching and the dashed lines stand for different levels of the 
search cost.  For instance, if the search cost is given by 
$c_1$, the solid and the dashed lines intersect twice and each 
intersection represents an NE.  Observe that an NE with $q \in 
(0,1)$ exists for search costs smaller than $\overline{c}$.  
However, a stable NE exists only if the search cost is between 
$\underline{c}(k)$ and $\overline{c}$ (like in many models of 
simultaneous consumer search such as \cite{burdettjudd1983}, 
\cite{fershtmanfishman1992}, \cite{janssenmoraga2004}, 
\cite{atayev2019a}).  To see that, suppose that the actual 
search cost is given by $c_1$ so that there are two NEs.  Then, 
only the NE which corresponds to the higher share of searching 
consumers, i.e., higher $q$, is locally stable.  This is because 
if the actual share of searching consumers is higher (lower) 
than the equilibrium one, the expected benefit of searching is 
lower (higher) than the cost of doing so.  Therefore, consumers 
have incentive to search less (more) and the actual share of 
searching consumers converges to the equilibrium one.  Applying 
a similar argument, we can see that the NE that corresponds to 
the lower share of searching consumers is unstable.%
\footnote{There always exists a stable equilibrium without 
	search (see, \cite{galeotti2004}), known as Diamond paradox 
	due to \cite{diamond1971}.  In this equilibrium, there is no 
	trade.  However, we are not interested in this unrealistic 
	equilibrium.}

Regarding the question of interest, we state the main result in 
the next proposition.

\begin{proposition}\label{prop:simul}
	In a stable NE with $q \in (0,1)$, as $k\to \infty$ the 
	lower bound of the search cost interval $\underline{c}(k)$ 
	converges to zero, the share of searching consumers 
	approaches zero, while price dispersion remains and
	the expected price is higher than the production marginal cost.
\end{proposition}

The proof is in the appendix and the intuition is as follows.  
As consumers have more friends who can potentially share price 
information, buyers' incentive to acquire price quotes through 
costly search falls.  In the limit when $k\to \infty$ the share 
of searching consumers almost vanishes.  To understand why price 
dispersion persists, we first note that the level of price 
dispersion is given by $\eta$.  Observe that $\eta$ is a 
ratio of captive consumers to that of price-comparing 
consumers.  Thus, if $\eta$ either approaches zero or goes to 
infinity, price dispersion disappears.  In the appendix, we show 
that if $\eta \to 0$ or $\eta \to \infty$, the expected benefit 
of searching, which is given by the right-hand side of 
\eqref{eq:IC_simul}, converges to zero.  However, this violates 
\eqref{eq:IC_simul} and, hence, cannot be the case.  Finally, as 
firms do not price below the production marginal cost, the 
presence of price dispersion means that the expected price is 
higher than the marginal cost of production.

\section{Endogenous Link Formation}\label{s:link_form}

In reality, obtaining information via WOM is costly in a sense 
that one has to spend time to request information from friends.  
For instance, texting a message/writing an email to friends or 
posting a question on an online forum requires some time.  At 
the same time, the current state of technology allows us to send 
the same message (or email) to a group of friends simultaneously 
and any post on an online forum can be seen by all users of the 
forum.  We address this issue in 
this section.  Particularly, we assume that each consumer can 
request price information from all her $k$ links by incurring 
cost $l\geq 0$.  Link formation is a binary decision: either a 
consumer forms $k$ links or does not form any links.  The link 
is directed, meaning that if consumer $i$ forms a link with 
consumer $j$, information flows from $j$ to $i$ only.  Let 
$\omega$ be the probability that a consumer invests in forming 
$k$ number of personal links. The rest of the model is the same 
as that in the previous section.  

Timing of the game is as follows.  First, firms simultaneously 
set prices. Second, without knowing prices, consumers 
simultaneously decide whether to form links and how many firms 
to search.  After search is complete, price information (if any) 
flows according to link formation. A consumer who observes at 
least one price can make a purchase.  We  use Nash equilibrium 
(NE) as a solution concept, and focus on symmetric equilibria. 

To derive an equilibrium, we first suppose that all consumers 
establish $k$ links.  Conditional on that assumption, we find 
consumers' optimal search strategy and firms' optimal pricing 
policy.  We know, however, that if all consumers establish $k$ 
links, in equilibrium consumers randomize between not searching 
and searching one firm for $c \in [\underline{c}(k), 
\overline{c}]$ and firms price according to \eqref{eq:F_simul} 
(recall Proposition \ref{prop:simul_eq}).  

Next, given these search and pricing strategies, we determine 
conditions under which consumers indeed prefer establishing $k$ 
links.  It is easy to check that 
a searching consumer prefers to form $k$ links if
\begin{equation}\label{eq:link_c1}
\left(1 - \left(1 - \frac{q}{2}\right)^k\right)(E[p]-E[\min\{p_1,p_2\}]) >l.
\end{equation}
Similarly, a non-searching consumer finds it worthwhile to form links if
\begin{equation}\label{eq:link_c2}
\left(1 - \left(1-q\right)^k\right)(v-E[p]) + \left(1 + (1-q)^k - 2 \left(1 
- \frac{q}{2}\right)^k\right)(E[p]-E[\min\{p_1,p_2\}]) >l.
\end{equation}
In the appendix, we show that \eqref{eq:link_c2} is implied by 
\eqref{eq:link_c1}.  The intuition is that, for a given $k$, the 
probability that a searching consumer receives the second price 
quote via WOM is lower than the probability that a non-searching 
consumer receives any price information from friends.  
Therefore, if a searching consumer who observes one price finds 
it worthwhile to form links, a non-searching consumer who does 
not observe any price definitely prefers to form links.  We show 
that, in equilibrium, searching consumers certainly form links 
if the cost of doing so is sufficiently small.  We state this 
result in the following proposition.

\begin{proposition}\label{prop:simul_endo}
	For any $v>0$ and $k\geq 1$, there exists $[\underline{c}(k), 
	\overline{c}] \subset(0,v)$ and $\overline{l}(k)>0$ such that for $c\in 
	[\underline{c}(k), \overline{c}]$ and $l\leq \overline{l}(k)$, there exists 
	a stable NE given by $\left(F(p), q,\omega \right)$, where 
	$F(p)$ is given 
	by \eqref{eq:F_simul}, $q$ is determined by \eqref{eq:IC_simul}, and 
	$\omega =1$.
	
	Furthermore, as $k \to \infty$, $\overline{l}(k)$ remains 
	strictly positive, the share of searching consumers 
	converges to zero, and price dispersion persists with the 
	expected price being higher than the production marginal 
	cost.
\end{proposition}

The proof is in the appendix.  Notice that the equilibrium in 
the proposition has similar characteristics as that in the 
previous section.  The only 
difference is the condition on $l$.  We can think of the model 
in Section \ref{s:simultaneous} as a special case of the current 
model where $l=0$ (and $k$ is fixed).  The reason why the 
equilibrium exists for sufficiently small $l$ is 
straightforward: if it was too costly to establish 
links, consumers would not form any links. 

For the limiting case of $k \to \infty$, the NE inherits the 
properties of that in Proposition \ref{prop:simul}, namely price 
dispersion remains and the firms retain some market power.  The 
intuition behind it is similar to that in the previous section, 
if consumers' incentive to form links does not decrease with 
$k$.   Consumers' incentive to form links increases with $k$, as 
the probability of obtaining price information via WOM 
increases.  As a result, an increase in $k$ does not mitigate 
consumers' incentive to form links and, hence, in the limit as 
$k\to \infty$, prices are dispersed just like in the previous 
section.

\section{Conclusion}\label{s:conclusion}

In the paper, we have revisited the role of costly information 
acquisition by consumers and its dissemination among consumers 
via WOM.  Specifically, we showed that, when the number of 
consumers' personal links increases without limits, firms earn 
positive profits and price dispersion persists if searching for 
prices is costly.  Also the result holds even when we allow for 
endogenous personal link formation.

We need to note that we considered a specific type of link 
formation: a consumer either forms $k$ links or does not form 
any links.  From a theoretical point, such modeling is 
restrictive as it implicitly assumes the following type 
of cost structure of link formation.  The cost of forming one 
link is positive and that of forming additional links (up to $k$ 
number) is free.  However, the cost of forming a link in 
addition to $k$ links is prohibitively high. A different way of 
modeling link formation is to assume linear or convex cost of 
link formation, which is common in the existing literature 
(e.g., \cite{galeottigoyal2010}).  Yet this introduces 
difficulties as, generically, a symmetric equilibrium does not 
exist because searching consumers have less incentive to form 
links than non-searching consumers.  As a result, undertaking 
comparative static analysis becomes difficult. This is certainly 
an area for future research.

\pagebreak
\appendix
\setcounter{equation}{0}
\renewcommand{\theequation}{\thesection.\arabic{equation}}

\section{Proofs}

\subsection{Proofs of Proposition \ref{prop:simul_eq} and \ref{prop:simul}}

For the proof of Proposition \ref{prop:simul_eq}, we mainly summarize the proof 
provided in \cite{galeotti2004}.  This will be helpful for us for the proof of 
the Proposition \ref{prop:simul}.  

First, we note some facts that we will use in the proof: $E[p] = 
v - \int_{\frac{\eta v}{1+\eta}}^{v}F(p)dp = v\eta \ln 
\left(1+\frac{1}{\eta}\right)$ and $E[\min\{p_1,p_2\}] = v-2 \int_{\frac{\eta 
v}{1+\eta}}^{v}F(p)dp + \int_{\frac{\eta v}{1+\eta}}^{v}[F(p)]^2dp$ so that 
\begin{equation*}
E[p] - E[\min\{p_1,p_2\}] = \eta v\left((1+2\eta)\ln\left(1 + 
\frac{1}{\eta}\right)-2\right).
\end{equation*}
Also, it is easy to check that $\lim\limits_{q \downarrow 0}\eta = \infty$ and 
$\lim\limits_{q \uparrow 1}\eta =\frac{1}{2(2^k-1)}$.

Second, we refer to \cite{galeotti2004} which shows that
\begin{equation*}
\begin{aligned}
\underline{c}(k) &&=&&& \lim\limits_{q \uparrow 1} \left\{(1-q)^k(v-E[p]) + 
\left(\left(1-\frac{q}{2}\right)^k - (1-q)^k\right)\big(E[p] - 
E[\min\{p_1,p_2\}]\big)\right\}\\
&&=&&& \frac{v}{2^k(2^k-1)}\left(\frac{2^{k-1}\ln\left(2^{k+1} 
-1\right)}{2^k-1}-1\right),
\end{aligned}
\end{equation*}
where it is easy to check that $0<\underline{c}(k)<v$.  The author also shows 
that that the derivative of the RHS of \eqref{eq:IC_simul} w.r.t. $q$ evaluated 
at $q \uparrow 1$ is negative.  From that facts that  $0<\underline{c}(k)<v$ 
and the RHS of \eqref{eq:IC_simul} is decreasing in $q$ in the neighborhood of 
$q \to 1$, there exists a stable NE for values of $c$ that is 
slightly above 
$\underline{c}(k)$.  This completes the proof Proposition \ref{prop:simul_eq}.

Now, we analyze the limiting case for the proof of Proposition 
\ref{prop:simul}.  It is clear that 
\begin{equation*}
\begin{aligned}
\lim\limits_{k \to \infty} \underline{c}(k) &&=&&& v 
\lim\limits_{k \to \infty} \frac{\ln(1+2(2^k-1))}{2(2^k-1)^2} - \lim\limits_{k 
\to \infty} \frac{v}{2^k(2^k-1)} = v\lim\limits_{x \to \infty} 
\frac{\ln(1+2x)}{2x^2}\\
&& \myeq&&& \lim\limits_{x \to \infty} \frac{1}{2x(1+2x)}=0,
\end{aligned}
\end{equation*}
where $x = 2^k-1$.

Next, we need show that $q \to 0$ when $k \to \infty$ by contradiction.  
Suppose that $\lim\limits_{k \to \infty}q >0$.  It implies that $\lim\limits_{k 
\to \infty}\eta = 0$.   Then, however, we have the RHS of \eqref{eq:IC_simul} as
\begin{equation*}
\Resize{}{
v\lim\limits_{k \to \infty}\left[(1-q)^k\left(1 - \eta 
\ln\left(\frac{1}{\eta}+1\right)\right) + \left(\left(1-\frac{q}{2}\right)^k 
-(1-q)^k\right)\eta\left((1+2\eta)\ln\left(1+\frac{1}{\eta}\right) 
-2\right)\right]=0,
}
\end{equation*}
since
\begin{equation*}
\begin{aligned}
\lim\limits_{k\to\infty}\eta\ln\left(\frac{1}{\eta}+1\right) &=&& 
\lim\limits_{\eta\to 0}\eta\ln\left(\frac{1}{\eta}+1\right)\\
&=&&\lim\limits_{z\to \infty}\frac{\ln(z+1)}{z}\\
&\myeq && \lim\limits_{z\to \infty}\frac{1}{z+1}\\
& = && 0,
\end{aligned}
\end{equation*}
\begin{equation*}
\begin{aligned}
\lim\limits_{k\to\infty}\eta\left(1 - 
\eta\ln\left(\frac{1}{\eta}+1\right)\right) &=&& 
\lim\limits_{\eta\to 0}\eta\left(1 - 
\eta\ln\left(\frac{1}{\eta}+1\right)\right)\\
&=&&\lim\limits_{z\to \infty}\frac{1 - \frac{\ln(z+1)}{z}}{z} = 
\lim\limits_{z\to \infty}\frac{z - \ln(z+1)}{z^2}\\
&\myeq && \lim\limits_{z\to \infty}\frac{1 - \frac{1}{z+1}}{2z} = 
\lim\limits_{z\to 
\infty}\frac{z}{2z(z+1)}\\
& = && 0,
\end{aligned}
\end{equation*}
and
\begin{equation*}
\begin{aligned}
\lim\limits_{k\to\infty}\eta\left((1+2\eta)\ln\left(1+\frac{1}{\eta}\right) 
-2\right) &=&& 
\lim\limits_{\eta\to 0}\left[\eta\ln\left(1+\frac{1}{\eta}\right) - 2 
\eta\left(1 - \eta\ln\left(\frac{1}{\eta}+1\right)\right)\right]\\
&=&&0 - 2\times 0 = 0,
\end{aligned}
\end{equation*}
where $z = 1/\eta$.  Thus, \eqref{eq:IC_simul} is violated, a contradiction.  
Hence, must be $\lim\limits_{k \to \infty}q=0.$

Finally, we show that $\eta$ remains strictly positive and finite in order to 
prove that the price dispersion remains when $k \to \infty$.  We will prove it 
by contradiction that $\eta$ cannot go to $0$ or $\infty$ when $k \to \infty$.  
The former case have been proven in the previous paragraph, and to prove the 
latter we suppose that $\lim\limits_{k \to \infty} \eta =\infty$.  Note that 
this implies that $0<\lim\limits_{k\to \infty}(1-q)^k \leq 1$ and  
$0<\lim\limits_{k\to \infty}\left(1-\frac{q}{2}\right)^k \leq 1$.  Then, we 
can evaluate the RHS of the indifference condition as
\begin{equation*}
\Resize{}{
	v\lim\limits_{k \to \infty}\left[(1-q)^k\left(1 - \eta 
	\ln\left(\frac{1}{\eta}+1\right)\right) + 
	\left(\left(1-\frac{q}{2}\right)^k 
	-(1-q)^k\right)\eta\left((1+2\eta)\ln\left(1+\frac{1}{\eta}\right) 
	-2\right)\right]=0,
}
\end{equation*}
since
\begin{equation*}
\begin{aligned}
\lim\limits_{k\to\infty}\eta\ln\left(\frac{1}{\eta}+1\right) &=&& 
\lim\limits_{\eta\to \infty}\eta\ln\left(\frac{1}{\eta}+1\right)\\
&=&&\lim\limits_{z\to 0}\frac{\ln(z+1)}{z}\\
&\myeq && \lim\limits_{z\to 0}\frac{1}{z+1}\\
& = && 1,
\end{aligned}
\end{equation*}
and
\begin{equation*}
\begin{aligned}
\lim\limits_{k\to\infty}\eta\left(1 - 
\eta\ln\left(\frac{1}{\eta}+1\right)\right) &=&& 
\lim\limits_{\eta\to \infty}\eta\left(1 - 
\eta\ln\left(\frac{1}{\eta}+1\right)\right)\\
&=&&\lim\limits_{z\to 0}\frac{1 - \frac{\ln(z+1)}{z}}{z} = 
\lim\limits_{z\to 0}\frac{z - \ln(z+1)}{z^2}\\
&\myeq && \lim\limits_{z\to 0}\frac{1 - \frac{1}{z+1}}{2z} = 
\lim\limits_{z\to 0}\frac{z}{2z(z+1)}\\
& = && \frac{1}{2},
\end{aligned}
\end{equation*}
and
\begin{equation*}
\begin{aligned}
\lim\limits_{k\to\infty}\eta\left((1+2\eta)\ln\left(1+\frac{1}{\eta}\right) 
-2\right) &=&& 
\lim\limits_{\eta\to \infty}\left[\eta\ln\left(1+\frac{1}{\eta}\right) - 2 
\eta\left(1 - \eta\ln\left(\frac{1}{\eta}+1\right)\right)\right]\\
&=&&1 - 2\left(\frac{1}{2}\right) = 0.
\end{aligned}
\end{equation*}
Again the indifference condition is violated, which proves that it cannot be 
$\lim\limits_{k\to\infty}\eta = \infty$.  Hence, $\eta$ must be strictly 
positive and finite when $k \to \infty$, which completes the proof.

\subsection{Proof of Proposition \ref{prop:simul_endo}}

For the proof, we first assume that all consumers form $k$ links and derive 
conditions under which consumers randomize over searching one firm and not 
searching. Later, we verify whether consumers indeed form $k$ links, given 
their search strategies.

We know from Section \ref{s:simultaneous} that if consumers have $k$ links, 
they randomize over searching one firm and not searching in a stable 
equilibrium for $c \in [\underline{c}(k),\overline{c}]$ (see Proposition 
\ref{prop:simul}).  Also firms' pricing policy is given by \eqref{eq:F_simul} 
and consumers' search strategy by \eqref{eq:IC_simul} and $q_0+q_1=1.$

Next, we determine conditions under which consumers choose to form links.  
We note that a searching consumer strictly prefers to form a link if
\begin{equation*}
v - E[p] + \left(1 - \left(1 - 
\frac{q}{2}\right)^k\right)(E[p]-E[\min\{p_1,p_2\}]) - c-l>v-E[p]-c.
\end{equation*}
The LHS of the equation represents a searching consumer's payoff that forming 
$k$ links yields and the RHS that not forming any links yields.  The inequality 
can be simplified to \eqref{eq:link_c1}. It is easy to see that a non-searching 
consumer strictly prefers to establish links if \eqref{eq:link_c2} holds.

Now, we show that \eqref{eq:link_c2} is implied 
by \eqref{eq:link_c1}, which means that we can ignore \eqref{eq:link_c2} from 
our further analysis.  It is easy to see that this is the case if the LHS of 
\eqref{eq:link_c1} is less than the LHS of \eqref{eq:link_c2}.  This is the 
case if
\begin{equation*}
\left(1 - \left(1-q\right)^k\right)(v-E[p]) > \left(\left(1 - 
\frac{q}{2}\right)^k- (1-q)^k\right)(E[p]-E[\min\{p_1,p_2\}]). 
\end{equation*}
As $1 - \left(1-q\right)^k> \left(1 - 
\frac{q}{2}\right)^k- (1-q)^k$ for any $k\geq 1$ and $q \in (0,1)$, the 
inequality certainly holds if the following is true:
\begin{equation*}
v- E[p] > E[p]-E[\min\{p_1,p_2\}].
\end{equation*}
We use the facts applied in the proof of Proposition \ref{prop:simul} to 
rewrite the inequality as
\begin{equation*}
1 - \eta \ln\left(1 + \frac{1}{\eta}\right) > \eta\left((1+2\eta)\ln\left(1 + 
\frac{1}{\eta}\right)-2\right), 
\end{equation*}
or
\begin{equation*}
\ln\left(1 + \frac{1}{\eta}\right)< \frac{1+2\eta}{2\eta(1+\eta)}.
\end{equation*}
As both the LHS and the RHS of the inequality go to $\infty$ when $\eta \to 0$ 
and both sides converge to $0$ when $\eta \to \infty$, the inequality holds if 
the derivative of the LHS is less negative than that of the RHS.  The 
derivative of the LHS is $-\frac{1}{\eta(1+\eta)} = - 
\frac{2\eta^2+2\eta}{2\eta^2(1+\eta)^2}$ and that of the RHS is $- 
\frac{2\eta^2 + 2\eta +1}{2\eta^2(1+\eta)^2}$.  The former is less negative 
than the latter for $\eta >0$, which means that the inequality holds.  This 
proves that the condition in \eqref{eq:link_c2} is implied by 
\eqref{eq:link_c1}.

Then, the cutoff value of forming links, denoted by $\overline{l}(k)$, solves 
\begin{equation*}
\left(1 - \left(1 - \frac{q}{2}\right)^k\right)(E[p]-E[\min\{p_1,p_2\}]) =l.
\end{equation*}
The LHS is positive for any $q \in (0,1)$ and independent of $l$.  Since the 
LHS is strictly increasing in $l$, there exists a unique solution to the 
equation.  This establishes the existence and uniqueness of 
$\overline{l}(k)>0$.

Finally, for the limiting case of $k \to \infty$, we provide the proofs as 
follows.  We, first, assume that in equilibrium all consumers form links and 
show that $q$ converges to $0$ and price dispersion remains  with $E[p]$ being 
higher than zero as $k \to \infty$.  Then, we demonstrate that consumers indeed 
prefer forming links.  Assume that consumers always form links.  Then, from the 
proof of Proposition \ref{prop:simul}, we know that $\lim_{k \to \infty}q=0$ 
and $0<\lim\limits_{k 
\to \infty} \eta<\infty$.  The latter means that the price dispersion remains, 
which, along with the fact that firms do not price below production marginal 
cost, implies that $\lim\limits_{k \to\infty}E[p]>0$. 

Now, we show that, given consumers' search behavior and firms' pricing policies 
as above,  consumers indeed prefer forming links as $k \to \infty$.  
Notice that this is the case if $\lim\limits_{k \to \infty}\overline{l}(k)>0$, 
or if the LHS of \eqref{eq:link_c1} is positive and does not converge to zero 
as $k\to \infty$.  As $\lim\limits_{k \to \infty}[1-(1-q/2)^k]\geq 0$ 
and $\lim_{k \to \infty}(E[p]-E[\min\{p_1,p_2\}]>0$ (or price dispersion 
remains), the LHS of \eqref{eq:link_c1} is positive.  It does not converge to 
zero if $\lim\limits_{k \to \infty}(1-q/2)^k <1.$  We prove the last inequality 
by contradiction.  Suppose $\lim\limits_{k \to \infty}(1-q/2)^k =1.$  Then, it 
must be that $\lim\limits_{k \to \infty}(1-q)^k =1$ as otherwise it means that
\begin{equation*}
\lim\limits_{k\to \infty}\eta = \lim\limits_{k \to \infty} 
\frac{\left(1-\frac{q}{2}\right)^{k+1} - (1-q)^{k+1}}{1 + (1-q)^{k+1} - 
	2 \left(1-\frac{q}{2}\right)^{k+1}} =  \frac{1 - \lim\limits_{k \to \infty}
	(1-q)^{k+1}}{-1+ \lim\limits_{k \to \infty} (1-q)^{k+1}} =-1,
\end{equation*}
which cannot be the case.  However, that $\lim\limits_{k \to \infty}(1-q/2)^k 
=1$ and  $\lim\limits_{k \to \infty}(1-q)^k =1$ also mean that
\begin{equation*}
\begin{aligned}
\lim\limits_{k\to \infty}\eta &&=&&& \lim\limits_{k \to \infty} 
\frac{\left(1-\frac{q}{2}\right)^{k+1} - (1-q)^{k+1}}{1 + (1-q)^{k+1} - 
	2 \left(1-\frac{q}{2}\right)^{k+1}}\\
 &&\myeq&&&  \lim\limits_{k \to \infty} 
 \frac{\left(1-\frac{q}{2}\right)^{k+1}\left[\ln\left(1 - \frac{q}{2}\right) - 
 \frac{k q'}{2(1-\frac{q}{2})}\right] - (1-q)^{k+1}\left[\ln\left(1 -  q\right) 
 - \frac{k q'}{(1-q)}\right]}{(1-q)^{k+1}\left[\ln\left(1 -  q\right) - \frac{k 
 q'}{(1-q)}\right] - 	2 \left(1-\frac{q}{2}\right)^{k+1}\left[\ln\left(1 - 
 \frac{q}{2}\right) - \frac{k q'}{2(1-\frac{q}{2})}\right]}\\
&&=&&& \lim\limits_{k \to \infty} \frac{-\frac{k q'}{2} + k q'}{-kq' +kq'} = 
\infty,
\end{aligned}
\end{equation*}
a contradiction.  Thus, it must be that $\lim\limits_{k \to \infty}(1-q/2)^k 
<1$. Then, the facts that $\lim_{k \to \infty}(E[p]-E[\min\{p_1,p_2\}]>0$ and 
$\lim\limits_{k \to \infty}(1-q/2)^k <1$ prove that the LHS of 
\eqref{eq:link_c1} is strictly positive as $k \to \infty$, or that 
$\lim\limits_{k \to \infty}\overline{l}(k)>0$. 

The proof of the proposition is complete.

\pagebreak 
\bibliographystyle{aer}
\bibliography{xx}{}

\ifx\undefined\bysame
\newcommand{\bysame}{\leavevmode\hbox 
to\leftmargin{\hrulefill\,\,}}
\fi

\end{document}